# Bayesian considerations on the multiverse explanation of cosmic fine-tuning

V. Palonen[1]

*Department of Physics, P.O. Box 43, FI-00014 University of Helsinki, Finland*
(Dated: September 3rd, 2009)

Short title: Bayesian considerations on the multiverse

The fundamental laws and constants of our universe seem to be finely tuned for life. The various multiverse hypotheses are popular explanations for the fine tuning. This paper reviews the four main suggestions on inference in the presence of possible multiple universes and observer selection effects. Basic identities from probability theory and previously unnoticed conditional dependencies of the propositions involved are used to decide among the alternatives.

In the case of cosmic fine-tuning, information about the observation is not independent of the hypothesis. It follows that the observation should be used as data when comparing hypotheses. Hence, approaches that use the observation only as background information are incorrect. It is also shown that in some cases the self-sampling assumption by Bostrom leads to probabilities greater than one, leaving the approach inconsistent. The "some universe" (SU) approach is found wanting. Several reasons are given on why the "this universe" (TU) approach seems to be correct. Lastly, the converse selection effect by White is clarified by showing formally that the converse condition leads to SU and its absence to TU. The overall result is that, because multiverse hypotheses do not predict the fine-tuning for *this* universe any better than a single universe hypothesis, the multiverse hypotheses fail as explanations for cosmic fine-tuning. Conversely, the fine-tuning data does not support the multiverse hypotheses.

---

[1] email: vesa.palonen@helsinki.fi

# Introduction

The fine-tuning of our universe's fundamental laws and constants [1] [2] [3] [4] [5] [6] requires an explanation [2] [7] [8]. For example, it seems that the strength of gravity has to be fine-tuned to an accuracy of $1/10^{36}$ [4] for life to exist. Depending on the researcher, the number of recognized fine-tuning constraints posed by life varies from some tens to a hundred.

A popular way to explain the happy coincidences are the varied multiverse hypotheses. Namely, if there are sufficiently many universes, it will not be surprising to find at least one universe with fine-tuned constants. Because we cannot observe our own non-existence, the observer selection effect (OSE) is thought to 'filter out' all the universes where observers can not exist. Indeed, many have taken the combination of multiverse theories with the observer selection effect as sufficient to explain fine-tuning [2] [9] [10].

Several inferences in physics have been made using anthropic constraints. Given that observers (we) exist, one can use this information and constrain the estimates on physical parameters or make predictions about them. This represent inference within a physical hypotheses and both the hypothesis and observability are used as background information, basically as constraints. This kind of inference should not be confused with the goal of the present paper which is inference concerning the overarching hypotheses themselves. As will be seen, when comparing hypotheses, the information about observability enters the comparison as data, not merely as a constraint.

A good reminder of problems with the use of observer selection effects is given by Leslie [2]: A prisoner being executed by a firing squad finds that all of the marksmen have missed and against all odds he is alive. The surviving prisoner now notes that one can only observe one's own survival and uses the observer selection effect to infer that his survival was just as likely to be due to chance than due to design. There is nothing to explain, claims the prisoner. In fact, any ridiculous hypothesis with a nonzero probability for survival is an equally good explanation. But this seems intuitively false.

The above paradox is even more visible when we modify the example a bit. Let us consider a case where the prisoner is blindfolded and cannot hear anything. The prisoner only knows that he has just survived an execution by a firing squad. Furthermore, the prisoner knows that there are two squads around, one with two marksmen, the other with a thousand marksmen. Which firing squad is more probable to have carried out the execution? If the prisoner uses the selection effect as advocated e.g. by E. Sober [10], he will infer that both squads are equally probable. Again, this kind of inference seems intuitively false. It seems that at least in some cases hypotheses which give a low probability for the observation (survival) should also be given a low probability in the hypothesis comparison.

This paper reviews and evaluates four popular suggestions on how to use the observer selection effect in the case of fine-tuning. These are the "assume the observation (AO)" approach above by Sober, the "self-sampling assumption (SSA)" approach by Bostrom [11], the "some universe (SU)" approach by e.g. Manson and Thrush [12], and the "this universe (TU)" approach by e.g. White [13] and Dowe [14]. Among these, mainly the SSA has been claimed to be a general theory for selection effects. AO and SSA are discussed first and reasons are given for why they are likely to be incorrect. The discussion then proceeds to SU and TU and reasons are given for why cosmic fine-tuning seems to be a TU case.

The main argument in this paper concerns the proper use of the probabilities. The precise shape of the probability distributions or the numerical values are not central to the argument. The argument remains essentially the same for all reasonable distributions. Ways of estimating some of the underlying distributions exist [15], but uniform probabilities are used here for clarity. Although some shortcuts would be available, the equations are mostly written down in detail because in my opinion in the end this is the clearest way of carrying out the inference.

## Principles of Bayesian hypothesis comparison

In this section the basic equations for Bayesian hypothesis testing are reviewed. Readers familiar with Bayesian methods may want to skip this section.

Probabilities will be used in the epistemic sense, denoting plausibilities first and frequencies of occurrence only as a result [16] [17]. It is to be understood that all probabilities are conditional on some background information $I$. $I$ will mostly be shown explicitly in the equations as this may help the reader to see the difference between its use and the use of other propositions in the calculation.

Bayes' theorem can be derived from the product rule of probability, which states that the probability for $A$ and $B$ can be written as

$$p(A \wedge B) = p(A|B)p(B), \qquad (1)$$

where by $A \wedge B$ we mean a logical conjunction ($A$ and $B$ are true) and $p(A|B)$ denotes a conditional probability of $A$ being true *given that* B is true. Alternatively one can also write

$$p(A \wedge B) = p(B|A)p(A). \qquad (2)$$

Equating the right sides of eqs. (1) and (2) one obtains the Bayes' theorem

$$p(B|A) = \frac{p(B)p(A|B)}{p(A)}. \qquad (3)$$

Eq. (3) is a general tool for probabilistic inversion; If the probability of $A$ given $B$, $p(A|B)$, is known, we can calculate the probability of $B$ given $A$, $p(B|A)$. However, this presupposes that we can estimate $p(B)$ and $p(A)$, the prior probabilities of $B$ and $A$.

Let us assume that $C_i$, $i = 1,...,n$ are a complete set of propositions. The *law of total probability* states

$$p(A) = \sum_i p(A \wedge C_i). \qquad (4)$$

Hence, if we have the joint probability density for $A$ and $C_i$, we can sum over all the possible $C_i$ to get the probability for $A$. This process of eliminating propositions by summing is called marginalization.

Let $H_n$ be some hypothesis and $D$ some measured data. Bayes' theorem (eq. (3)) gives the probability of the hypothesis $H_n$ given the data $D$

$$p(H_n | D \wedge I) = \frac{p(H_n | I) p(D | H_n \wedge I)}{p(D | I)}, \qquad (5)$$

where $p(D | H_n \wedge I)$ is the probability of the data for the hypothesis $H_n$ (often called the likelihood) and $p(H_n | I)$ is the prior probability of $H_n$. The term $p(D | I)$ is often called a marginal probability and can be viewed as just a normalization constant common to all hypotheses.

Using eq. (5), we can compare the probabilities of several hypotheses by using the ratio of their probabilities given the data

$$\frac{p(H_n | D \wedge I)}{p(H_m | D \wedge I)} = \frac{p(H_n | I)}{p(H_m | I)} \frac{p(D | H_n \wedge I)}{p(D | H_m \wedge I)}, \qquad (6)$$

where on the right side the first term is the ratio of the prior probabilities of the hypotheses. The second term is the ratio of probabilities for the measured data $D$ under each hypothesis. It is therefore a ratio of each model's prediction of the data $D$.

In order to clarify the use of the above equation, Figure 1. depicts the predictions of two hypotheses for the variable $E$. It is important to note that probability distributions are normalized so that the total probability, which is a sum or an integral over all possible values, equals one. In the figure, hypothesis $H_1$ gives a rather broad prediction and due to normalization, the prediction is low overall. Hypothesis $H_2$ gives a rather strict prediction for $E$ near to the value 2 and the distribution is high in this area. Now, if we measure $E = 2$, $p(E = 2 | H_1 \wedge I)$ has a moderate value but $p(E = 2 | H_2 \& I)$ is large. From eq. (6) we get $\frac{p(H_1 | E = 2 \wedge I)}{p(H_2 | E = 2 \wedge I)} = \frac{0.35}{1.6} \ll 1$ and hence hypothesis $H_2$ is more probable given the data. However, if we measure $E = 3$, which does not hit the high peak of $H_2$, $\frac{p(H_1 | E = 3 \wedge I)}{p(H_2 | E = 3 \wedge I)} = \frac{0.35}{0.001} \gg 1$ and hence hypothesis $H_1$ is more probable. It is seen that because of the normalization, hypotheses are penalized for making broad predictions. They give a moderate probability for many possible values but cannot predict anything well.

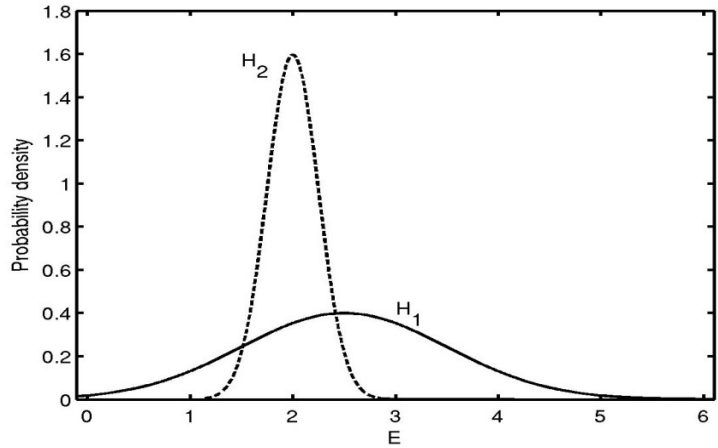

**Figure 1. Two hypotheses H1 and H2 have a prediction concerning the value of E. Due to normalization of the probability density, a hypothesis with a broad prediction does not predict anything very well.**

In addition to the above equations, some tools for handling the conditional independence of statements, originally discovered in the field of Bayesian belief networks, will be used. They will be introduced below.

# Nomenclature

The following symbols will be used frequently, the reader may wish to skim casually through them for now and refer back to the definitions in detail later if necessary:

$H_n$     The proposition that hypothesis $H_n$ is true.

$M$     Multiverse hypothesis in particular

$\Omega$     The set conceivable values for the vector of physical constants

$E$     A vector of the coordinate variables of corresponding to the physical constants and parameters.

$D$     Our present data (concerning a realization of $E$), $D \leftrightarrow$ "$E$ = the present physical constants and parameters."

Later in the paper we will need to distinguish between "this universe is fine-tuned" and "some universe is fine-tuned". For this reason the following two are introduced:

$D_t$     This universe is fine-tuned. The index $t$ represents the 'true index' of our universe.

$D'$     Some universe is fine-tuned

And we will also be using the following:

$\Omega_{E,n}$     A set of all possible values for $E$, given hypothesis $H_n$

$\Omega_{O,n}$     A set of all values of $E$ that can be observed given hypothesis $H_n$

$\Omega_O$   A set of a all values of $E$ that can be observed, an union of all $\Omega_{O,n}$, $\Omega_O \leftrightarrow \bigcup_n \Omega_{O,n}$

$O$   A proposition "The realization of $E$ can be observed", $O \leftrightarrow \bigcup_n H_n \wedge (E \in \Omega_{O,n})$

$S(\Omega)$   The size (cardinality) of $\Omega$

For clarity, we will call a universe inside a multiverse a subverse. Our universe, which is possibly one subverse, will be called ourverse.

## Assume the observation (AO)

E. Sober [10] and Ikeda and Jefferys [18] have argued that because one cannot observe one's own non-existence, the observability $O$ should be taken as background information for hypothesis comparison. If this is the case, using the Bayes' theorem to obtain the probability of hypothesis $H_n$ given the fine-tuning data $D$, one would get

$$\begin{aligned} p(H_n | D \wedge O \wedge I) &= p(H_n | I) p(D | H_n \wedge O \wedge I) / p(D | I) \\ &\propto p(D | H_n \wedge O \wedge I) \\ &= p(D | H_n \wedge E \in \Omega_{O,n} \wedge I) \\ &\approx \frac{1}{S(\Omega_{O,n})}, \end{aligned} \qquad (7)$$

where we have neglected $p(H_n | I)$, the prior probability for the hypothesis, and $p(D | I)$, the marginal probability in the second line. The prior and marginal probabilities are roughly the same for all hypothesis and will roughly cancel out when $H_n$ is compared to other hypotheses. The remaining term $p(D | H_n \wedge O \wedge I)$ is the prediction of the hypothesis for the data $D$. The important thing to note in the above result is that $O$ is used as a condition for the probabilities in the same way as the background information $I$ is. This filters out all unobservable events and is equivalent to assuming that all hypotheses will only produce observable events. The prediction of fine-tuning for each hypothesis is therefore improved to certainty in the AO approach. This is a bit like asking for the probability of a six in a dice throw knowing that you will get a six. Hence, assuming AO and taking a uniform prediction for the cosmic constants in the last line of eq. (7), the probability of a hypothesis in the comparison is only dependent on the size of the observation-space.

Note that usually in Bayesian hypothesis comparison hypotheses are penalized for making broad predictions because the prediction is a normalized probability distribution. Yet, on AO, the non-observable part of the prediction is filtered out and the hypotheses are not penalized for predicting however broadly in the event-space.

Weisberg [19] has noted that our information in the case of observer selection effects is actually of the form "*If we observe, we will observe $O$*" not "*we will observe $O$*", as AO assumes. So, we really have only the conditional "$D \to O$" as background information. Weisberg shows that this information does not raise the prediction for fine-tuning.

Already the above argument seems enough to discredit AO but we will nevertheless press on to facilitate further understanding on the topic by using some of the mathematics developed for inference with Bayesian belief networks [20] [21]. In general, the probability for the hypothesis given the relevant information will depend on the joint probability, as can be seen from the definition of conditional probability

$$p(H_n | D \wedge O \wedge I) = p(D \wedge O \wedge H_n \wedge I) / p(D \wedge O \wedge I). \quad (8)$$

Now, the joint probability will in general factor as

$$p(D \wedge O \wedge H_n \wedge I) = p(I) p(H_n | I) p(O | H_n \wedge I) p(D | O \wedge H_n \wedge I). \quad (9)$$

Only in the case that the observation $O$ and the hypothesis $H_n$ are independent, $O \perp H_n | \varnothing$ (or $O \perp H_n | I$) [20] [21], does one get

$$\begin{aligned} p(D \wedge O \wedge H_n \wedge I) &= p(I) p(H_n | I) p(O | H_n \wedge I) p(D | O \wedge H_n \wedge I) \\ &= p(I) p(H_n | I) p(O | I) p(D | O \wedge H_n \wedge I) \\ &\propto p(H_n | I) p(D | O \wedge H_n \wedge I), \end{aligned} \quad (10)$$

which, when used in eq. (8), is the AO method. In the last line we show only the terms dependent on the hypothesis and hence relevant for hypothesis comparison. Hence, a necessary and sufficient condition for Sober's method is the independence of the observation from the hypothesis. Figure 1a shows the necessary probabilistic dependency structure as a graph [22] [23].

In the case of cosmic fine-tuning, the probability of observation is *not* independent of the hypothesis but instead the observation is a probabilistic product of the hypothesis, and one has to use eq. (9). The case corresponds to the graph in Figure 1b. Hence, in the case of cosmic fine-tuning, probability theory requires one to use also the prediction for the observation, $p(O | H_n \wedge I)$, in the hypothesis comparison. It's use is also required by the strong condition to use all available information. As a result, hypotheses purporting to explain fine-tuning must be penalized for having a low prediction for observability. So, when one has survived a very dangerous accident, the capability to observe ones own survival cannot be taken as background information, thereby assuming that the survival was certain, but as part of the data. Hypotheses must be favored according to how high a probability they give for the survival and the corresponding observation. The conditional statement "$D \to O$" by Weisberg can legitimately be taken as background information but this does not raise the prediction for fine-tuning.

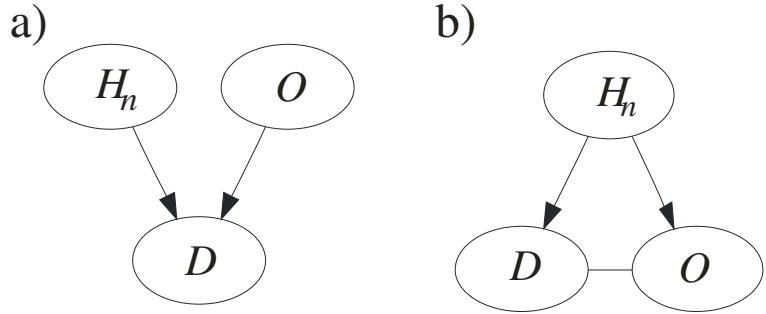

**Figure 2. a)** When the observation $O$ is independent of the hypothesis $H_n$ conditional on the data $D$, the observation cannot be used as data in the hypothesis comparison, resulting in a result equivalent to the assume the observation (AO) approach. **b)** In the case of cosmic fine-tuning the observation is a product of the hypothesis and hence $O$ has to be used as data in the hypothesis comparison. It follows that the AO approach is incorrect for cosmic fine-tuning.

## The self-sampling assumption (SSA)

Interesting work has been done by N. Bostrom [11] on developing a theory of how to take selection effects into account, along with a good explanation of the issues involved. Bostrom's suggestion for inference in cases involving selection effects is the self-sampling assumption (SSA), which states that one should reason as if one were a random sample from the set of all observers in one's own reference class. The exact definition of 'one's own reference class' remains open, leaving room for movement under criticism [24]. Bostrom has developed a modified version of SSA, called the observation equation or SSSA-R (from Strong SSA Revised), which mainly corrects for SSA's bias to select hypotheses with big amounts of observers. SSA was developed inductively from examples and from attempted solutions to some encountered paradoxes. This makes the method somewhat ad-hoc. It is based on intuitive solutions, which may or may not be truly correct solutions, to thought-experiments, which may or may not be analogous to fine-tuning. I my opinion SSSA-R and variants try to give directly something that should be the result of rigorous probabilistic inference.

Leslie has argued that SSA leads to observer-relative chances [25]. Bostrom has admitted that it does [26], but does not think that this poses a problem. Yet it does seem problematic that, even if two observers exchange information so that they have the same information except that the one has 'I am A, he is B' and the other has 'I am B, he is A', they will not arrive at the same probabilities with SSA. If they start gambling, both cannot win. Moreover, in principle the observers could 'rise above' their circumstances and, like we, calculate the probability that the other would use and hence both of them will obtain two contradictory probabilities for the case. A further question is, what probability should we use as outsiders, and why should the observers involved not use that probability also? To me it does not seem rational that the probability will change just because of the identity of the person doing the inference. Indeed, this is contrary to the starting assumptions of some derivations of probability theory [16] which may lead to contradictions in SSA since probability theory was used in the derivation of SSA and because SSA purports to give out probabilities.

We will now show that, in addition to the above problems, SSSA-R can lead to probabilities bigger than one. We will use Bostrom's God's Coin Toss (GC) example in reference [11] and modify it a bit: A coin is thrown. If the coin lands heads (H), a world with an amount $h_\beta$ of black bearded and $h_\alpha$ red bearded observers is created. If the coin lands tails (T), a world with $t_\beta$ black bearded and zero red bearded observers is created. Let $p(H) = p(T) = 1/2$ and let $D_\beta$ represent the evidence available to the observer who has discovered that he has a black beard. Now, with Bostrom's own usage of SSSA-R (the original approach where all observers are in the same reference class), we get

$$p(H \mid D_\beta) = \left( \frac{1}{h_\alpha + h_\beta} + \frac{1}{t_\beta} \right)^{-1} \frac{1}{h_\alpha + h_\beta}. \qquad (11)$$

and

$$p(T \mid D_\beta) = \left( \frac{1}{h_\alpha + h_\beta} + \frac{1}{t_\beta} \right)^{-1} \frac{1}{t_\beta}. \qquad (12)$$

But a problem becomes apparent when we calculate

$$p(T) = p(T \wedge D_\beta) + p(T \wedge \neg D_\beta) = p(T \wedge D_\beta) = p(T \mid D_\beta) p(D_\beta), \qquad (13)$$

and hence

$$p(D_\beta) = \frac{p(T)}{p(T \mid D_\beta)}. \qquad (14)$$

Now because $p(T) = 1/2$ and $p(T \mid D_\beta)$ can be arbitrarily small, SSSA-R can lead to arbitrarily large values for $p(D_\beta)$. For example, taking $h_\beta = h_\alpha = 2$ and $t_\beta = 10$, we get $p(D_\beta) = \frac{1/2}{2/7} = \frac{7}{4}$. But by definition a probability cannot be bigger than one. A similar result can be obtained for SSA. Hence, SSA and variants are inconsistent with probability theory.

## This universe (TU) or some universe (SU)?

We will now turn our attention on whether one should really use $D_t \Leftrightarrow$ "this universe is fine-tuned" (TU) or $D' \Leftrightarrow$ ="some universe is fine-tuned" (SU) as data. Of course, this choice has profound implications on the inference concerning fine-tuning because in a sufficiently large multiverse it is virtually certain that some subverse will be fine-tuned.

White has argued for the use of TU as data in the fine-tuning case [13]. White points out that some-propositions are weaker and contain less information than corresponding this-propositions. Hence, White argues that because the some-proposition $D'$ contains less information than the this-proposition $D_t$, which we have available, $D_t$ should not be replaced with $D'$ because doing so would violate the strong desiderata to use all available information. So, when a this-proposition is available as data, the use of the corresponding some-proposition is warranted if and only if it leads to the same conclusion as using the this-proposition. We will first try to answer the main criticisms of TU and then show why cosmic fine-tuning is indeed a TU case.

Manson and Thrush argue against the use of TU on the basis that those arguing for TU do not seem to use "this planet" (TP) as data [12]. This criticism is dependent on two assumptions:

(1) We should not use TP as data.

(2) Subverses and planets are equivalent as concerns the inference.

Consider a simple case where we did not have any observations of the other planets. Let us also say we had two hypotheses about solar system formation. The good hypothesis $H_1$ would predict the observed properties of the earth and the sun with good accuracy and the bad hypothesis $H_2$ would give a near-zero prediction for the data. Science normally proceeds by selecting the hypothesis which better explains the phenomena, in this case $H_1$. Yet, if we were to use "some planet is such and such" (SP) as data, even the bad hypothesis would predict that the data will be observed somewhere with a probability of one when there is an infine amount of planets. The two hypotheses would remain equal. It therefore seems that the past, present, and future success of astrophysics depends on our steering clear of SP.

Second, there is a difference in inference concerning planets and universes: we can observe other planets but not other subverses. So, for planetary formation, our data consists also of the other observed planets and hypotheses are compared based on all of the planets. In the case of the multiverse, we *do not* have data on the other universes and therefore our inference will be based only on this universe. As will be shown below, this difference is important when considering whether one should use this- or some-type propositions. Hence, subverses and planets are not analogous for the present discussion.

Bostrom's analysis of fine-tuning also uses SU. Bostrom argues against TU [11] first by pointing out that White's approach is committed to $p(M|D' \wedge D_i) < p(M|D')$ for every subverse $i \neq t$. The above means that the probability for a multiverse is smaller when TU is used as data than when SU is used as data. So far so good, but next Bostrom misinterprets the inequality to mean that any finely tuned universe other than ourverse would make $M$ more likely, which he then sees as a problem for TU. But what makes $p(M|D')$ bigger is not the state of any particular universe but instead the relaxation of the information that we have about the particular universe. A small modification of White's coffee table example [13] shows that this is a natural inequality resulting from the difference between 'some' and 'this': The information that someone at the coffee table was drunk last night increases the probability for there being several persons at the table. Yet the use the information that in fact the person who was drunk last night was John cancels the increase of probability for there being several people at the table.

Second, Bostrom tries to prove wrong White's assertion that a multiverse does not predict fine-tuning for this universe any better than a single universe by pointing out that there may be correlations among subverses. Bostrom seems to shift the burden of proof to White, claiming that unless it is proved that a multiverse cannot gain a better prediction for fine-tuning due to correlations, a multiverse is a good explanation. But possible correlations to other subverses cannot improve a multiverse's prediction for fine-tuning because we do not possess relevant information about the other possible subverses. Neither do we have sufficient knowledge about the correlation function. Using epistemic probabilities we can marginalize over our ignorance and as the sum is taken over the possible states of the other subverses, the prediction about ourverse will of course iron out irrespective of the correlations. And marginalization over the possible correlation functions would have the same effect. Hence, correlations among the subverses do not help the multiverse.

Third, Bostrom goes on to claim that White is committed to denying the above-mentioned inequality $p(M | D' \wedge D_i) < p(M | D')$ because, in Bostrom's view, $D'$ can not be relevant to $M$ unless information about one subverse $x$ can be probabilistically relevant to another subverse $y$. But this is clearly wrong. Even if the events of a hypothesis are independent, compound information about the events can be very relevant. It is hard to think of a case where this would not hold. It seems that while White's view is not committed to paradoxes, Bostrom's criticism is, for, were Bostrom's criticism of TU true, we would all be forced to use "some" in almost every inference, and as a result would be committing the gambler's fallacy perpetually. It seems that none of the criticisms of TU carry much weight.

In addition to reasons given by White, other points can be made in support of TU:
- We do seem to live in this universe. Our uncertainty about ourverse concerns the 'true index' or 'true label' of ourverse, not the ourverse itself.
- The use of TU is needed if one wants to make any inference in a large multiverse. This is because, in a sufficiently large multiverse, almost everything happens somewhere. It is curious that multiverse hypotheses have been used rather exclusively as explanations for fine-tuning. Using SU would lead to the predictions of almost all theories being very nearly unity for almost all conceivable data. This would end scientific inference.
- A TU analysis will not change drastically should we discover the true index or true 'name' of ourverse. Yet this information seems to be purely indexical since for the present purposes we already know that the index is something (call it $t$).
- A TU analysis seems to minimize the expectation value of observers being wrong.

To summarize, there are good arguments for using TU, no good reasons against it, and good reasons against SU. This seems sufficient to establish the use of TU for cosmic fine-tuning. Yet, one can further advance the point by considering more carefully the available information.

## TU and the converse conditional

Presently we know that we live in this universe but we do not know the true index or the 'name' of ourverse. Is this a problem for TU? Note that if we do not know the true index of the ourverse, it follows that we do not know the true index of anything within it. For example, for any given dice throw in the ourverse there may be corresponding dice throws in other subverses. If the correct inference would be to use 'some event in the multiverse' in these cases, no inference could be done. Clearly, probability theory works mostly fine when we use "this", so we can expect that our not knowing the true index of ourverse will not be a problem for TU. Let us now look at how the unknown true index can result in "this" or "some", depending on the case at hand.

As before, let $D_t$ be the proposition that this universe is finely tuned for life and let also $t$ be a variable for the true index of ourverse. Let the number of subverses be $N$. It is important to note that what is uncertain here is the indexing or naming, not the subverse itself. That is, the observed fine-tuning will not move to another subverse because of a change in our knowledge, just the indexing we are using may move. Now, because we do not know the value of $t$, we will express our ignorance by marginalizing (summing) over it:

$$p(\bigcup_{i=1}^{N}(t=i) \wedge D_t \mid H) = p(\bigcup_{i=1}^{N}(t=i \wedge D_t) \mid H)$$
$$= \sum_{i=1}^{N} p(t=i \wedge D_t \mid H) = \sum_{i=1}^{N} p(D_t \mid H) p(t=i \mid D_t \wedge H) \qquad (15)$$
$$= \sum_{i=1}^{N} p(D_t \mid H) \frac{1}{N} = p(D_t \mid H).$$

The above result is equivalent to TU. Note that our knowledge about $D_t$ does not affect the probability about the true index of ourverse. For all we know, ourverse's index could be any index and hence the principle of indifference suggests a uniform prior for $p(t=i \mid D_t \wedge H)$. It follows that because cosmic fine-tuning is a TU case, a multiverse does not predict fine-tuning any better than a single universe and conversely fine-tuning does not support a multiverse.

White points out that what makes the difference between "this" and "some" seems to be what White calls the converse observational selection effect [13]. This refers to whether any event happening entails our observing it, e.g. whether, using the notation above, $D_i \to t=i$ is true. We will denote this conditional by $S_c \leftrightarrow 'D_i \to t=i'$. Of course, in the case of cosmic fine-tuning, we will not observe every universe that is fine-tuned, instead we observe ourverse only. Hence $S_c$ does not hold for cosmic fine-tuning. However, in cases where the converse conditional $S_c$ is true and we observe one $D_t$, we will typically get

$$p(\bigcup_{i=1}^{N}\left(t=i \wedge D_t \wedge \bigcap_{j \neq i} \neg D_j\right) | H \wedge S_c)$$

$$= \sum_{i=1}^{N} p(t=i \wedge D_i \wedge \bigcap_{j \neq i} \neg D_j | H \wedge S_c)$$

$$= \sum_{i=1}^{N} p(D_i | H) p(t=i | D_i \wedge H \wedge S_c) \prod_{j \neq i} p(\neg D_j | H) \qquad (16)$$

$$= \sum_{i=1}^{N} p(D_i | H) p(t=i | t=i \wedge H) \prod_{j \neq i} p(\neg D_j | H),$$

$$= N p(D_t | H) p(\neg D_\alpha | H)^{N-1}$$

which is a "some"-type result. It can also be shown that in the more general case where we observe $m$ events and given that we would observe all events for which $D_\alpha$ is true (converse condition), the likelihood will be of the form

$$p(D_\alpha | H)^m p(\neg D_\alpha | H)^{N-m} \frac{N!}{(N-m)! m!}, \qquad (17)$$

which clearly is a "some"-type result. Hence, the above clarifies why the converse conditional leads to "some" and why, when the converse conditional does not hold, as is the case with cosmic fine-tuning, the correct approach is "this" analysis. As an example, a fisherman fishing with a fishnet and capable of seeing all the fish couth should use a "some"-type analysis like the one in eq. (17). Yet a fish, if it cannot see other fish, should use a "this" analysis like the one in eq. (15). The thing which makes the difference is information the observer has, not the identity.

## Conclusions

The four most viable approaches for inference in a possible multiverse and in the presence of an observer selection effect were reviewed.

Concerning the 'assume the observation' (AO) approach advocated by Sober, Ikeda, and Jefferys, it was shown that this kind of an observer selection effect is justified if and only if the observation is conditionally independent of the hypothesis. In the case of cosmic fine-tuning the observation would be a child of the hypothesis and the two are not independent. It follows that one should use the observation as data and not as a background condition. Hence, the AO approach for cosmic fine-tuning is incorrect.

The self-sampling assumption approach by Bostrom was shown to be inconsistent with probability theory.

Several reasons were then given for favoring the 'this universe' (TU) approach and main criticisms against TU were answered. A formal argument for TU was given based on our present knowledge. The main result is that even under a multiverse we should use the proposition "*this* universe is fine-tuned" as data, even if we do not know the 'true index' of our universe. It follows that because multiverse hypotheses do not predict fine-tuning for this particular universe any better than a single universe hypothesis, multiverse hypotheses are not adequate explanations for fine-tuning. Conversely, our data on cosmic fine-tuning does not lend support to the multiverse hypotheses. For physics in general, irrespective of whether there really is a multiverse or not, the common-sense result of the above discussion is that we should prefer those theories which best predict (for this or any universe) the phenomena we observe in our universe.

## Bibliography

[1] J. Barrow and F. Tipler, *The Anthropic Cosmological Principle*, Oxford University Press, 1988.

[2] J. Leslie, *Universes*, Routledge, 1990.

[3] M. Denton, *Nature's Destiny: How the Laws of Biology Reveal Purpose in the Universe*, Free Press, 2002.

[4] R. Collins, "Evidence for fine-tuning", in *God and Design*, Routledge, 2003, pp. 178-199.

[5] G. Gonzalez and J. Richards, *The Privileged Planet*, Regnery Publishing, 2004.

[6] M. Rees, *Just Six Numbers*, Basic Books, 2001.

[7] R. Swinburne, *The Existence of God, 2nd ed.,* Oxford University Press, 2004.

[8] R. Collins, "The Teleological Argument", in *The Blackwell Compation to Natural Theology, 2009*.

[9] S. Weinberg, "Living in the Multiverse", in *Universe or Multiverse?,* Cambridge University Press, 2007, arXiv:hep-th/0511037.

[10] E. Sober, "The Design Argument", in *The Blackwell Guide to the Philosophy of Religion*, Blackwell Publishing, 2004.

[11] N. Bostrom, *Observation Selections Effects and Probability, Doctoral dissertation*, London School of Economics. Available at anthropic-principle.com, 2000.

[12] N. A. Manson and M. J. Thrush, "Fine-tuning, Multiple Universes, and the "This Universe" Objection", *Pacific Philosophical Quarterly 84*, pp. 67-83, 2003.

[13] R. White, "Fine-Tuning and Multiple Universes", *Nous 34*, p. 260–76, 2000.

[14] P. Dowe, "Response to Holder: Multiple Universe Explanations are not Explanations", *Science and Christian Belief 11*, pp. 67-68, 1999.

[15] J. Koperski, "Should We Care about Fine-Tuning?", *British Journal for the Philosophy of Science 56(2)*, pp. 303-319, 2005.

[16] E. T. Jaynes, *Probability Theory: The Logic of Science*, Cambridge University Press, 2003.

[17] D. J. C. MacKay, *Information Theory, Inference, and Learning Algorithms*, Cambridge University Press, available at http://www.inference.phy.cam.ac.uk/mackay/itila/book.html, 2002.

[18] M. Ikeda and W. H. Jefferys, "The Anthropic Principle Does Not Support Supernaturalism", in *The Improbability of God*, Prometheus Press, 2006, pp. 150-166.

[19] J. Weisberg, "Firing Squads and Fine Tuning: Sober on the Design Argument", *British Journal for the Philosophy of Science 56(4)*, 2005.

[20] J. Pearl, "Bayesian Networks", *UCLA Cognitive Systems Laboratory, Technical Report (R-246)*, http://ftp.cs.ucla.edu/pub/stat_ser/R246.pdf.

[21] A. P. Dawid, "Influence Diagrams for Causal Modelling and Inference", *Intern. Statist. Rev. 70*, pp. 161–189, http://www.ucl.ac.uk/Stats/research/Resrprts/abs01.html#221, 2002.

[22] J. Whittaker, *Graphical Models in Applied Multivariate Statistics*, John Wiley & Sons, 1990.

[23] R. E. Neapolitan, *Probabilistic Reasoning in Expert Systems*, John Wiley & Sons, 1990.

[24] K. D. Olum, "Conflict Between Anthropic Reasoning and Observation", arXiv:gr-qc/0303070v2, 2003.

[25] J. Leslie, "Observer-relative Chances and the Doomsday Argument", *Inquiry 40*, pp. 427-436, 1997.

[26] N. Bostrom, "Observer-relative Chances in Anthropic Reasoning?", *Erkenntnis 52*, pp. 93-108, 2000.